\documentclass[twocolumn,showpacs,aps,prl,amsmath,amssymb,floatfix,%
superscriptaddress]{revtex4}
\usepackage{graphicx}% Include figure files
\usepackage{epsfig}
\usepackage{dcolumn}% Align table columns on decimal point
\usepackage{bm}% bold math
\begin{document}
%\wideabs{
\title{Vibration-induced Kondo tunneling through metal-organic
  complexes with even electron occupation number}
\author{K. Kikoin}
\affiliation{Physics Department, Ben-Gurion University of the
Negev, Beer-Sheva 84105, Israel}
\author{M.N. Kiselev}
\affiliation{Institute f\"ur Theoretische Physik, Universit\"at
W\"urzburg, 97074 W\"urzburg, Germany}
\author{M.R. Wegewijs}
\affiliation{Institut f\"ur Theoretische Physik - Lehrstuhl A,
RWTH Aachen, 52056 Aachen, Germany}
\date{\today}
\begin{abstract}
We investigate transport through a mononuclear transition-metal
complex with strong tunnel coupling to two electrodes. The ground
state of this molecule is a singlet while the first excited state
is a triplet. We show that a modulation of the tunnel-barrier due to
a molecular distortion which couples to
 the tunneling \emph{induces} a Kondo-effect, provided
the discrete vibrational energy compensates the singlet/triplet gap.
We discuss the single-phonon and two-phonon assisted co-tunneling
 and possible experimental realization of the theory.
\end{abstract}
\maketitle
\begin{figure}[h]
  \includegraphics[width=7cm,angle=0]{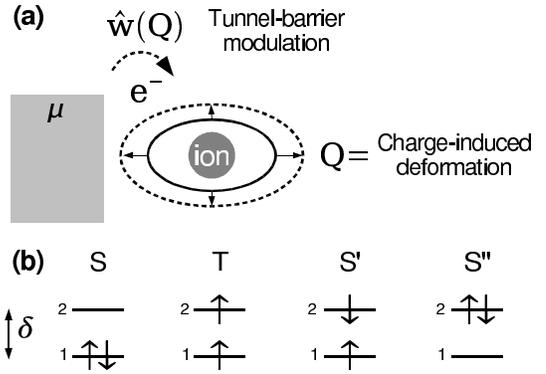}
  \caption{Schematic situation:
    (a) Electrode tunnel-coupled to a transition-metal organic complex.
    Charging of the complex by a tunnel process deforms the
    outer part of the ligand cage
    without strongly affecting the direct
    coordination-sphere of the metal ion and thereby the ligand-field splitting.
    We assume that the extra electron is localized mainly
    on the cage.
    Electrons tunnel onto the ion through the tails of the
    molecular state centered on the ion, which includes admixtures of the
    outer shell electronic states.
    Therefore
    the main effect of the charging is the \emph{modulation of the
      tunnel barrier} between the ion-centered states and electrode.
    (b) Electronic $e$-type states discussed in the text in order of increasing
    energy.
  }\label{fig:scheme}
\end{figure}

 The effect of quantized vibrational motion of a
single-molecule device on the electron transport has been
investigated intensely. Vibrational effects have been observed in
the sequential tunneling regime~\citep{Park00,Pasupathy04,LeRoy04}
as well as the strong tunneling
regime~\citep{Park02,Yu04inel,vanderZant06}. Theoretical attention
first focused on the weak coupling limit, where exchange of
mechanical energy quanta with tunneling electrons (vibration
assisted
tunneling)~\citep{Boese01,McCarthy03,Braig03a,Mitra04b,Koch04b,Wegewijs05}
and modulation of the tunnel barriers
(shuttling)~\cite{Fedorets03,McCarthy03} were discussed. In the
strong tunneling limit the effect of vibrations on the
Kondo-anomaly in the linear conductance was discussed due to
assisted tunneling~\citep{Paaske04,Cornaglia04,Cornaglia05} and
also due to tunnel-barrier modulation~\citep{Cornaglia05}. In this
Letter, however, we demonstrate that discrete vibrations through
tunnel-barrier modulation can~\emph{induce} a Kondo effect in a
transition-metal (TM) organic complex (TMOC) with \emph{even}
number of electrons by compensating for the singlet-triplet
splitting {\em at zero bias}. Already for molecules of moderate
size, many vibrational modes are available, underlining the
importance of considering the above effect. Our effect is
essentially different from the phonon-assisted Kondo effect
in~\citep{Cornaglia04,Cornaglia05}: no strong electron-vibration
coupling is required
 nor special gate-voltage restrictions.
\\
We study the transport through a TMOC with a TM ion secluded in a
ligand cage. The cage is in tunnel contact with metallic
reservoirs (surface, STM nanotip, or edges of metallic wire in
electro-migration or break junction geometry). Fig.~\ref{fig:scheme}(a) illustrates this setup.
 We consider a TMOC with even electron number $N$ fixed by
charge and energy quantization.
% Coulomb blockade.
 The ground state is supposed to be  a spin singlet, and
 the energy of the lowest  triplet excitation $\Delta$ exceeds  Kondo temperature
 $T_K$. The linear conductance is thus suppressed.
 To investigate how intramolecular vibrations may {\em
induce} transport through a Kondo effect, in the first place one
should incorporate a vibronic mode in a generic tunneling
Hamiltonian
%%% PREVIOUS WORKS
\begin{eqnarray}\label{ham}
  H=H_{mol}+H_{res}+ H_{tun}
\end{eqnarray}
Here $H_{mol}$ includes the $3d$ electron levels in a ligand field
of the cage electrons, the molecular orbitals of these ligands, as
well as interactions within the $3d$ shell and within the cage.
One should take into account the three most relevant charge states
including their dependence on the vibrational coordinate of the
cage $Q$:
\begin{equation}\label{Hmol}
  H_{mol}=H^{(N)}_Q+H^{(N+1)}_Q+H^{(N-1)}_Q+T_{n}
\end{equation}
The last term $T_n$ is the kinetic energy of the cage distortion.
The eigenstates of $H^{(N \pm 1)}_Q$  are admixed to those of $H^{(N)}$ by the tunneling
$H_{tun}$ of electrons from the reservoir $H_{res}$:
\begin{equation}\label{res}
H_{res}+H_{tun}=\sum_{k\sigma} \epsilon_{k}
c^\dag_{k\sigma}c_{k\sigma} + \hat w_Q\sum_{k\mu\sigma}\left(\tilde
d^\dag_{\mu\sigma}c_{k\sigma}+ H.c.\right) .
\end{equation}
Effectively a single electrode remains after the standard rotation of
electron states~\cite{Glazman88}.
We assume that there is no direct tunneling contact between the 3d
electrons of the caged TM and the electrodes. However the $3d$-orbitals are
hybridized with ligand molecular orbitals, and tunneling
becomes possible due to small overlap between the "tails" of
distorted orbitals $\tilde d$ and the lead electrons (see caption
to Fig.~\ref{fig:scheme}(a) for more details).

We write $H_{mol}$ for
the low-lying many particle states of the TMOC
 in the compact form
  \begin{eqnarray}\label{dot}
  H^{(N)}_{mol}
  &=&  \sum_{\Lambda=S,T0,T\pm} E_{\Lambda}(Q) X^{\Lambda \Lambda}
  \nonumber
\end{eqnarray}
where $X^{\Lambda \Lambda'}=| \Lambda \rangle \langle \Lambda' |$
are so called Hubbard operators describing transitions between the
eigenstates $\Lambda,\Lambda'$ of the TMOC, and all many-particle
effects and their coordinate dependence are included in the
energies $E_{\Lambda}(Q)$. Similarly, states of the charged TMOC
with $N \to N\pm 1$ are denoted as $|\gamma\rangle$. The
microscopic origin of the low-energy singlet (S) and triplet (T)
is most easily conceived as follows. The evenly occupied ligand
cage is in a singlet spin state, and it does not affect the
structure of the spin multiplet. We assume that the ligand field
of the cage has low symmetry, so that the 5-fold orbital
degeneracy is completely lifted. For instance, for a distorted
tetrahedral symmetry of the ligand field we focus on the
configuration $d^2(e^2)$ with two $e$-states split by $\delta$ due
to this distortion. In case of distorted cubic symmetry, the same
S/T multiplet arises for the configuration $d^8(t^6e^2)$.
Fig.~\ref{fig:scheme}(b) illustrates the states of the TMOC: it is
seen that in the case of a weak intrashell exchange $I<\delta$ the
ground state is singlet S and the lowest excitation is the S/T
transition. The energy difference $\Delta\equiv E_T-E_S=\delta-I$
is assumed to be larger than the Kondo temperature $T_K$. The
excitation energies of two singlet excited states $S'$ and
$S^{''}$ are $\delta$ and $2\delta$, respectively. We ignore below
these singlet excitations, since they are not involved in the
Kondo tunneling.

We use the simplest approximation of single
harmonic vibration mode with frequency $\Omega$. We assume that
the relative shifts of the harmonic potentials in different
excited and charged states is negligibly small i.e. for all
$x=\Lambda,\gamma$ we have $E_{x}(Q)\approx E_{x}+ \Omega Q^2/2$ and
$T_n=\Omega P^2/2$. The coordinate $Q$ is normalized to the
zero-point motion. In this weakly non-adiabatic limit vibrations
are slow compared to the electron motion on the molecule. No
vibrational excitations can be induced by the tunneling in the
linear transport regime, unless the distortion of the ligand cage
$Q$ affects the tunnel amplitude $\hat{w}(Q)$. Then {\em local
phonons} can be emitted or absorbed. The tunneling Hamiltonian may
be rewritten in the form
\begin{equation}\label{tun}
  H_{tun}= \hat{w}(Q)
  \sum_k \sum_{\Lambda\gamma \sigma}^\prime \left[X^{\Lambda \gamma} c_{k\sigma}+
    H.c.\right]
\end{equation}
(here the second summation is restricted by spin selection rules).
This situation is opposite to the anti-adiabatic limit
 where the vibrations are fast and many phonon
excitations are involved in the Kondo exchange~\cite{Paaske04}.
\begin{figure}[h]
  \includegraphics[width=7cm,angle=0]{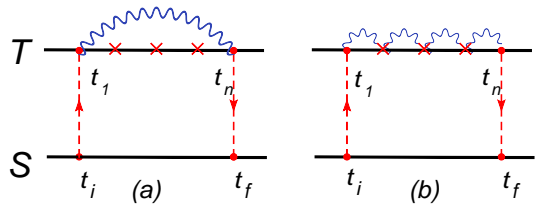}
  \caption{\label{fig:process}
  Two types of phonon-assisted Kondo co-tunneling processes.
    (a) Virtual  phonon absorption initiates a S/T transition, Kondo processes
    take place
    in an intermediate triplet state and the
    phonon is emitted in the end.  (b) Every spin flip process in the
    intermediate triplet state is accompanied by a two-phonon
    process. Points and crosses denote spin-flip process in the S/T
    and T/T channels, $t_{i,f}$ are initial and final times,
    and $t_1...t_n$ denote the intermediate co-tunneling acts.
  }
\end{figure}

Thus we consider the effect of the modulation of the tunnel
amplitude on the linear transport in the Coulomb blockade regime.
In this regime one has to eliminate the tunneling term from the
Hamiltonian by summing over virtual processes where $N\pm1$
electrons occupy the molecule and vibrational quanta are excited
in these virtual states. This procedure known as a
Schrieffer-Wolff (SW) transformation, leads us to a Kondo
Hamiltonian (Eq.~(\ref{hkon}) below) for a S/T multiplet
{\cite{Kikoin01}} and an oscillator with {\em electron-phonon
interaction built into the exchange coupling}. According to
Ref.~\cite{Kikoin01}, transitions within the singlet/triplet spin
manifold are described by two vectors ${\bf S}$ and ${\bf R}$,
where ${\bf S}$ is the usual spin 1  vector  and ${\bf R}$ is a
vector describing S/T transitions. These two vectors are
constructed by means of Hubbard operators $X^{\Lambda\Lambda'}$ in
the following way \cite{Kikoin01}:
\begin{eqnarray}\label{shub}
S^+ & = & \sqrt{2}\left(X^{10}+X^{0-1}\right),~ S^z =
X^{11}-X^{-1-1},
 \label{m.1} \\
R^+ & = & \sqrt{2}\left(X^{1S}-X^{S-1}\right), ~R^z =
-\left(X^{0S}+X^{S0}\right). \nonumber
\end{eqnarray}

\begin{figure}[h]
  \includegraphics[width=7cm,angle=0]{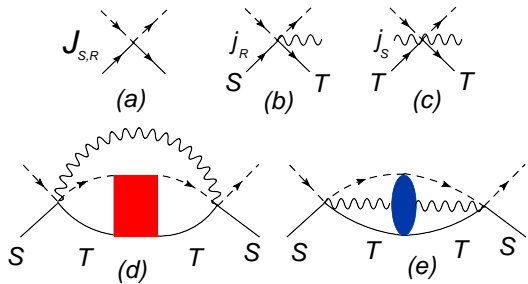}
  \caption{(a): Bare exchange vertices $J_{S,R}$;
  (b) Single phonon correction  $j_{R}$ to the vertex
    $J_{R}$; (c) Two-phonon correction $j_{S}$
    to the vertex $J_{S}$; (d,e) Renormalized vertices $\gamma_{1,2}$
    corresponding to the processes illustrated by Fig.
    \ref{fig:process}(a), (b), respectively.}\label{fig:vert1}
\end{figure}

The effective  Hamiltonian arising after the SW transformation has
the form
\begin{equation}\label{hkon}
{H}_{eff}=H_{res}+\frac{1}{2}\Delta {\bf S}^2 +
   {\hat J}_S {\bf S}\cdot {\bf s}
 + {\hat J}_R {\bf R}\cdot {\bf s}
 + \frac{\Omega}{2} P^2
 \end{equation}
The electron spin operator is given by the conventional expansion
$ {\bf
s}=\frac{1}{2}\sum_{kk'}\sum_{\sigma\sigma'}c^\dag_{k\sigma}{\mbox{\boldmath
$\tau$}}_{\sigma \sigma'} c_{k'\sigma'}$ where ${\mbox{\boldmath
$\tau$}}$ is a Pauli vector. The exchange coupling constants
${\hat J}_{S,R}(Q)$ are estimated as ${\hat J}_{S}(Q)\approx
\sum_\gamma|\hat w(Q)|^2/|E_T-E_\gamma|$, and ${\hat
J}_{R}=\alpha{\hat J}_{S}$. Here $\alpha<1$ is a coefficient
arising because of admixture of singlet states $S^\prime$,
$S^{\prime\prime}$ to the ground state \cite{Kikoin01}. In the
weakly non-adiabatic regime the kinetic energy in the denominator
has been neglected and the nearly identical $Q$ dependence are
practically cancelled out in the addition energies. Thus, no
multiphonon replicas appear in the denominators of ${\hat
J}_{S,R}$ unlike Ref. \cite{Paaske04}.
\\
The main source of phonon
emission/absorption in our case is the tunneling rate $|\hat
w(Q)|^2$.
Expanding it in the quantized displacement operator
$Q=(b^\dag+ b)/\sqrt{2}$ we come to phonon
assisted exchange vertices presented in Fig. \ref{fig:vert1}.
In accordance with Fig.
\ref{fig:process}, we retain only single-phonon processes for
$\hat J_R(Q)$ and only two-phonon processes for $\hat J_S(Q)$:
\begin{equation}
\hat J_S(Q)=J_ S+j_S Q^2,~~~~ \hat J_R(Q) = J_R + j_R Q.
\end{equation}
It is obvious that $j_S\ll j_R$.
To draw these vertices we used the fermionic representation for
the operators (\ref{shub})
\begin{eqnarray}
S^+ &=&  \sqrt{2}(f_0^\dagger f_{-1}+f^\dagger_{1}f_0),\;\;
%S^- =  \sqrt{2}(f^\dagger_{-1}f_0+ f_0^\dagger f_{1}) ,\;\;\;\;
S^z =
f^\dagger_{1}f_{1}-f^\dagger_{-1}f_{-1},\\
R^+  &=& \sqrt{2}(f^\dagger_{1} f_s -  f_s^\dagger f_{-1}),
\;\; R^z = -( f_0^\dagger f_s + f_s^\dagger f_0).\nonumber
\end{eqnarray}
The  spin-fermion,  electron and phonon
propagators are presented in Fig. \ref{fig:vert1} by solid, dashed
and wavy lines, respectively.
Physically, Figs~\ref{fig:process}(a) and \ref{fig:vert1}(d) describe the regime, where
the virtual phonon absorbed by the TMOC in the process of co-tunneling
brings the energy necessary to compensate the S/T gap $\Delta$.
Then the Kondo effect develops as multiple spin-flip co-tunneling
in the triplet state, and the system returns to the ground state
singlet after virtual phonon emission.
The vertex corrections are calculated by means of analytical
continuation of Matsubara-type diagrams from imaginary axis to the
real frequency axis.
Summation of all parquet
diagrams entering the 4-tail box in Fig. \ref{fig:vert1}(d) gives
for $\gamma_1$
\begin{equation}
\gamma_1 \sim (j_R)^2\rho
\left[\frac{\log\left(\displaystyle\frac{D}{{\rm
max}[T,|\Delta-\Omega|]}\right)} {1-J_S A
\rho\log\left(\displaystyle\frac{D}{{\rm
max}[T,|\Delta-\Omega|]}\right)}\right]
\end{equation}
here $A\sim 1$ is a constant determined by spin algebra. The Kondo
temperature extracted from this equation reads
\begin{equation}
T^{(1)}_K \sim D \exp\left(-\frac{1}{A\rho J_S}\right)
\end{equation}
Here $D$ is the effective width of the electron conduction band and
$\rho$ is the density of states on the Fermi level.
The second channel illustrated by Figs \ref{fig:process}(b) and
\ref{fig:vert1}(e) involves two single-phonon processes compensating
S/T transitions. The exchange acting in the intermediate triplet
states is accompanied by \emph{two}-phonon processes. Then summation of
parquet diagrams entering the 4-tail vertex in Fig.
\ref{fig:vert1}(e) gives for $\gamma_2$
\begin{equation}
\gamma_2 \sim (j_R)^2\rho
\left[\frac{\log\left(\displaystyle\frac{D}{{\rm
max}[T,|\Delta-\Omega|]}\right)} {1-j_S A'
\rho\log\left(\displaystyle\frac{D}{{\rm
max}[T,|\Delta-\Omega|]}\right)}\right]
\end{equation}
The Kondo temperature characterizing this channel depends on the
phonon-assisted exchange constant $j_S$
\begin{equation}
T^{(2)}_K \sim D \exp\left(-\frac{1}{A'\rho j_S}\right)\ll
T^{(1)}_K
\label{Tk}
\end{equation}
One concludes from these calculations that the single-phonon
processes are sufficient to compensate the energy of the S/T splitting
and induce resonance tunneling through the TMOC provided a local
vibration mode with appropriate frequency satisfying the condition
\begin{equation}\label{ineq}
|\Omega-\Delta|\lesssim T_K^{(1)}
\end{equation}
exists in the cage. One can expect in this case a significant
enhancement of the tunnel conductance already at $T>T^{(1)}_{K}$
according to the law $G/G_0 \sim \ln^{-2} (T/T^{(1)}_{K})$~\cite{Glazman88},
where $G_0$ is the conductance at unitarity limit $T \to 0$.

Thus we formulated in this paper conditions under which phonons
are not only involved in Kondo screening but even induce Kondo
tunneling. In is worth mentioning that in spite of the fact that
the Kondo effect exists in our case {\em only under phonon
assistance}, the Kondo temperature (\ref{Tk}) is \emph{the same} as in
the usual Kondo effect. Since $T_K$ is high enough ($\sim 10$ K)
in electro-migrated junction experiments with a TMOC deposited between
contacts~\cite{Park02,Yu04inel,vanderZant06},
 the effect predicted in this work seems to be easily
observable. The crucial point is the existence of phonon
satisfying condition (\ref{ineq}) in a TMOC with the S/T multiplet
as a lowest spin excitation. One should note, however, that even
if this condition is not exactly satisfied, one may tune the
system by applying the magnetic field. Then the triplet is split,
and only the level $E_{T\downarrow}=E_T-E_Z$ is involved in the
phonon induced Kondo tunneling ($E_Z$ is the Zeeman energy). Then
$\Delta$ in (\ref{ineq}) is substituted by $\Delta_Z=\Delta-E_Z$,
and $E_Z$ may be tuned until the inequality is satisfied.  Thus
the vibration gives rise to a magnetic field induced Kondo effect
at Zeeman energies which can be much smaller that $\Delta$. The
Kondo screening takes place due to the processes presented in Figs
\ref{fig:process}(a) and \ref{fig:vert1}(d). The only difference
is that in this case the effective spin of the TMOC is one half
instead of one \cite{Pust00}. The theory of this effect will be
presented in a separate publication.

Another way to tune the condition (\ref{ineq}) is to stretch the
break junction and thereby distort the TMOC and change the
frequency $\Omega$. Such mechanical control was demonstrated
experimentally for the H$_2$ molecule~\cite{Djukic05}.
There it was also shown that the isotope effect may be used
 for the same purpose.

The generic feature of the phonon-assisted Kondo screening
discussed in this work is that only the {\em virtual} phonon
excitations enter the co-tunneling amplitude to compensate the S/T
energy gap $\Delta$, so that the Kondo effect manifests itself as
a zero-bias anomaly (ZBA). Another mechanism of such
compensation was discussed in Ref. \cite{KK03}, where the energy
deficit was covered by the conduction electron acceleration at
finite bias. In that case, the Kondo regime arises in
non-equilibrium condition, so it is fragile against dephasing effects \cite{KK03,Rosch03}. In our case the system
remains in thermodynamic equilibrium around ZBA, so that the only
limiting factor of this sort is the lifetime of local vibration
mode, which is usually long enough in comparison with $\hbar/T_K$.

To conclude, we demonstrated in this work that phonon
emission/absorption can induce Kondo tunneling in a transition-metal
organic complex with even
electron occupation and a spin singlet ground state, when the
conventional Kondo effect is suppressed. Unlike the situation
studied in the current literature~\cite{Paaske04,Cornaglia04,Cornaglia05}, where the influence of real
phonon excitations on the conventional Kondo effect is discussed,
and various kinds of side-band satellites due to the polaronic
effect are considered, we appeal to virtual phonon excitation, so
that the system remains in a quasi-elastic tunneling regime. One of
essential ingredients of our theory is that we use the dynamical
symmetry of the TMOC, which characterizes both the spin algebra of
localized spin itself and transitions between various levels of
different spin multiplets \cite{Kikoin01}. In our case the dynamical symmetry
group is $SO(4)$.

Since the tunnel contact
between magnetic ion and metallic reservoir in TMOC is mediated by a
ligand cage, one may be sure that the relevant
vibration excitations are the local phonons characterizing this
cage. Although we confined ourselves to a specific model with
two electrons in an $e$ subshell, the mechanism is quite general. The
theory may be easily modified for any system with the same
structure of the lowest spin multiplets. One of such examples is the so
called "Fulde molecule"~\cite{Fuldemol} schematically representing the spectra
of lanthanocenes, where the rare-earth magnetic ion is sandwiched
between two rings of CH radicals. In that case the number of
electrons in the cage and in the $4f$-shell is {\it odd}. Another
candidate is the endofullerene family with atoms~\cite{NC60} or
magnetic ions within a carbon cage~\cite{LnC82}.
The theory of vibration-induced Kondo effect may be also generalized
for the case of degenerate modes  and for more
complicated spin-multiplets including half-integer spins. These
issues will be discussed in forthcoming publications.

This work is partially supported by the SFB-410 and ISF grants. MK
acknowledges support through the Heisenberg program of the DFG.
Part of this work is done during KK's stay in the Max
Planck Institute of Complex Systems, Dresden.
MRW acknowledges financial support through the
EU RTN Spintronics program HPRN-CT-2002-00302.

%\bibliography{cite1.bib}

\begin{thebibliography}{99}
\bibitem{Park00}H. Park, et al.,
% J. Park, A. K. L. Lim, E. H. Anderson, A. P.
%Alivisatos, and P. L. McEuen
Nature {\bf 407}, 52 (2000).
\bibitem{Pasupathy04}A. N. Pasupathy, et. al.,
% J. Park, C. Chang, A. V. Soldatov,
%S. Lebedkin, R. C. Bialczak, J. E. Grose, L. A. K. Donev, J. P.
%Sethna, D. C. Ralph,
Nano Lett. {\bf 5} (2005).
\bibitem{LeRoy04}B. J. LeRoy, S. G. Lemay, J. Kong, and C. Dekker, Nature 432, 371 (2004).
\bibitem{Park02}J. Park, et al.,
% A. N. Pasupathy, J. I. Goldsmith, C. Chang, Y. Yaish, J. R. Petta, M. Rinkoski,
%J. P. Sethna, H. D. Abruna, P. L. McEuen,
 Nature {\bf417}, 722 (2002).
\bibitem{Yu04inel}L. H. Yu, et al.,
%Z. K. Keane, J. W. Ciszek, L. Cheng, M. P. Stewart,
%J. M. Tour, and D. Natelson,
Phys. Rev. Lett. {\bf 93}, 266802 (2004);
L. H. Yu and D. Natelson, Nano Lett. {\bf 4}, 79 (2004).
\bibitem{vanderZant06}
H. S. J. van der Zant, et. al.,
%Y.-V. Kervennic, M. Poot, K. ONeill, Z. de Groot, J. M. Thijssen and
%H. B. Heersche, N. Stuhr-Hansen, T. Bj\"ornholm, D. Vanmaekelbergh,
%C. A. van Walree and L. W. Jenneskens
Faraday Discuss., available online, (to appear in 2006).
\bibitem{Boese01}D.  Boese and H.  Schoeller,  Eur.  Phys.  Lett.  {\bf 54},  668
(2001).
\bibitem{McCarthy03}K.  D.  McCarthy,  N.  Prokofev,  and M.  T.  Tuominen,
Phys. Rev. B {\bf 67}, 245415 (2003).
%\bibitem{Flensberg03}K. Flensberg, Phys. Rev. B {\bf 68}, 205324 (2003).
\bibitem{Braig03a}S. Braig and K. Flensberg,  Phys. Rev.  B {\bf 68},  205324
(2003).
\bibitem{Mitra04b}A. Mitra, I. Aleiner, and A. J. Millis, Phys. Rev. B {\bf 69},
245302 (2004).
\bibitem{Koch04b}J. Koch and F. von Oppen, Phys. Rev. Lett. {\bf 94}, 206804
(2005).
\bibitem{Wegewijs05}
M. R. Wegewijs and K. C. Nowack, in: {\it Focus on NEMS}, New J.
Phys. (online), Eds. R. Blick and M. Grifoni.
\bibitem{Fedorets03}D. Fedorets, Phys. Rev. B {\bf 68}, 033106 (2003);
D. Fedorets et al.,
% L. Y. Gorelik, R. I. Shekhter, and M. Jonson,
Phys. Rev. Lett. {\bf 92}, 166801 (2004).
\bibitem{Paaske04}J. Paaske and K. Flensberg, Phys. Rev. Lett. {\bf 94}, 176801
(2005).
\bibitem{Cornaglia04}P. S. Cornaglia, H. Ness, and D. R. Grempel, Phys. Rev.
Lett. {\bf 93}, 147201 (2004).
\bibitem{Cornaglia05}P. S. Cornaglia, D. R. Grempel and H. Ness, Phys. Rev.
B {\bf 71}, 075320 (2005).
\bibitem{Kikoin01} K. Kikoin and Y. Avishai,  Phys. Rev. Lett. {\bf 86}, 2090 (2001);
Phys.  Rev.  B  {\bf 65},   115329 (2002).
\bibitem{KK03} M.N. Kiselev, K. Kikoin and L.W. Molenkamp, Phys.
Rev. B {\bf 68}, 155323 (2003).
\bibitem{Rosch03} A. Rosch, J. Paaske, J. Kroha, and P.W\"olfle,
Phys. Rev. Lett. {\bf 90}, 076804 (2003).
\bibitem{Pust00} M. Pustilnik, Y. Avishai, and K. Kikoin,
Phys. Rev. Lett. {\bf 84}, 1756 (2001).
\bibitem{Fuldemol} C.-S. Neumann, P. Fulde,
Z. Phys. B, {\bf 74}, 277 (1989);
M. Dolg et al,
%P. Fulde, W. K\"uchle, C-S. Neumann and H. Stoll,
J. Chem. Phys. {\bf 94}, 3011 (1991).
\bibitem{Glazman88}
L. I. Glazman and M. E. Raikh,
JETP Lett. {\bf 47}, 452 (1988).
\bibitem{NC60}
T. Almeida Murphy et al,
%Th. Pawlik, A. Weidinger1, M. H\"ohne, R. Alcala, and J.-M. Spaeth,
Phys. Rev. Lett. {\bf 77}, 1075 (1996)
\bibitem{LnC82}
R. D. Johnson,  et al.,
%M. S. de Vries, J. Salem, D. S. Bethune, C. S. Yannoni,
Nature {\bf 355}, 239 (1992).
\bibitem{Djukic05}
D. Djukic, et al.,
%, K.S. Thygesen, C. Untiedt, R.H.M. Smit, K.W. Jacobsen and J.M. van Ruitenbeek,
Phys. Rev. {\bf B 71}, 161402(R) (2005).
\end{thebibliography}

\end{document}